\documentclass[12pt]{article}

\title{New variational principles in classical and semiclassical
mechanics}
\author{G. Karl \\
Department of Physics, University of Guelph, Ontario NIG 2W1,
Canada \\ email: gk@physics.uoguelph.ca \\
V.A. Novikov \\ ITEP,
Moscow, 117218, Russia \\ email: novikov@heron.itep.ru}
\date{}

\begin{document}
\maketitle

\begin{abstract}

We demonstrate that reciprocal Maupertuis' Principle is the
classical limit of Schr\"{o}dinger's Variational Principle in
Quantum Mechanics.

\end{abstract}

\vspace{5mm}

 Misha Marinov loved analytical mechanics and understood
its beauty. Canonical transformations, Poisson structures,
symplectic geometry etc. were the notions that he constantly
used in his original papers, and in his famous review on path
integral. One of the authors (NV) had the pleasure to attend his
remarkable lectures at ITEP that preceded the review paper. The subtle
relation between Quantum and Classical Mechanics was one of the
major subjects in these lectures.

 We believe that Misha would have enjoyed to read that there exists a
 new formulation of Classical Mechanics (new variational principle)
that follows
from  Quasiclassical limit of Quantum Mechanics. This paper is
based on the results we published in reference [1], and reference [2]
in collaboration with Chris Gray.

\section{Schr\"{o}dinger's Quantum Variational Principle and its Classical Limit.}
The fundamental theory is Quantum Mechanics.
Classical Mechanics is only a special limit of Quantum Mechanics.
In other words Quantum mechanics can be a starting point for the
derivation of Classical mechanics. One expects therefore to derive
the variational principles of classical mechanics from the variational
principles of quantum mechanics.

It is widely known that Hamilton's Principle can be derived
in the framework of Feynman path integral formulation of quantum
mechanics [3]. We use the old Schr\"{o}dinger version of quantum
mechanics to derive new principle of classical mechanics. In
some sense this principle can be considered as a reciprocal to the known
Maupertuis principle.

The details of reaching the limit of classical mechanics from the
quantum domain are notoriously delicate and difficult.
 We start by recalling that
Schr\"{o}dinger's variational principle of wave mechanics has the
form

\begin{equation}
\delta[<\psi_n|\hat H|\psi_n>/<\psi_n|\psi_n>] =0,\
\end{equation}
where the quantum system has a Hamiltonian  $\hat H (\hat p,\hat q)$ which is a
function of the operators  of  position $\hat q$,  and conjugate momenta $\hat p$.
 Although this
principle is employed most often to find the ground state $|\psi_0>$
of the quantum system, the principle applies to all eigenstates $|\psi_n>$ of
the operator $\hat H$  (see e.g. ref. [4]). This point is elaborated in detail in [1],
where more references can be found.

At large quantum numbers $n$ we step into the domain of Classical Mechanics.
We want to demonstrate that at
large quantum numbers $n$, the Schr\"{o}dinger principle turns
into the following classical variational principle
\begin{equation}
0 = \left(\delta\frac{<\psi|\hat H|\psi>}{<\psi|\psi>}\right)_n
\stackrel{n\gg 1}{\longrightarrow} \delta \left(\frac{1}{T} \int^T
_0 H(q, p)dt \right)_W \equiv (\delta \bar E)_W = 0 \;\; ,
\end{equation}
where mean energy is defined as
\begin{equation}
\bar E=\frac{1}{T} \int^T_0 H(q, p)dt ,
\end{equation}
and action $W$ is
\begin{equation}
W=\int^B_A p dq ,
\end{equation}
where $T$ is  the time of propagation of the system from initial
point $A$ to final point $B$
in configuration space.

We start with the simplest case of periodic motion
in one dimension.
On the RHS of (2) the Hamiltonian $H(q, p)$ is the classical
counterpart of the quantum Hamiltonian $\hat H (\hat p,\hat q)$ . On the
LHS of (2) the quantum variation is made at large $n$, with
$n$ fixed [1]. The state $|\psi_n(q)>$ corresponds to a classical
(periodic) trajectory with precisely the same energy $E_n$.
We consider the trial wavefunctions and trial trajectories such
that they match each other.

When we use a WKB representation for all trial wavefunctions on
the LHS of (2) we have

\begin{equation}
\int\psi^*\psi dq = {\rm const}
\int^{q_{\rm max}}_{q_{\rm min}} dq/v = {\rm const}\int^T_0 dt =
{\rm const}\; T,
\end{equation}
 where $v$ is the velocity and $T$ is the period
of motion. In the same approximation the numerator of the LHS
(2) becomes ${\rm const} \int^T_0 H(q, p) dt$, with the same
constant as the denominator. Therefore the quantum expectation on
the LHS becomes a classical time average on the RHS of (2), i.e.
\begin{equation}
\frac{<\psi|\hat H|\psi>}{<\psi|\psi>}{\longrightarrow} \frac{1}{T} \int^T
_0 H(q, p)dt  \equiv \bar E \;\; ,
\end{equation}
 if
we use WKB wavefunctions for $|\psi_n>$. This of course is well
known.

All that remains to be discussed is how the constraint of fixed
$W$ arises on the RHS of (2). We recall that quantization
means that only certain classical energies possible on the RHS
match quantum energies $E_n$ on the LHS. For periodic motion in
one dimension, the constraint on the energy for an allowed quantum
state $n$, derived in the WKB approximation, is for large $n$,
\begin{equation}
W = \oint p dq = 2\int^{q_{\rm max}}_{q_{\rm min}}
\sqrt{2m(E_n -V(q))} dq = nh \;\; .
\end{equation}
This is the Bohr-Sommerfeld-Wilson quantization rule, which for
$n$ fixed means precisely that the action $W$ is to be kept fixed
on the RHS of (2), if $n$ is fixed on the left.

Thus for one-dimensional system we have completed the derivation of the Classical Variational
Principle from Quantum Mechanics. It says that variation of mean energy $\bar E$
at fixed short action $W$ is zero for true path. It looks like the new variational
Principle different from the Hamilton and Maupertuis Principles.

Next step is to consider systems with
$N$ degrees of freedom. Unfortunately in this general case
Quantum Mechanics does not help much. For integrable systems we can prove that
quantum average can be replaced by time average over classical quasi-periodic
motions when all quantum numbers ${n_i\gg 1}$. For chaotic non-integrable
motions this is rather plausible statement for very large time $T$.
As for the reliable proof, it is absent yet. But the main trouble is with
the classical analog of quantum numbers $n$.
 We guess that for general systems it is classical action W
that has to replace quantum numbers $n$ in quasiclassical limit
( exactly like in the case of dimension one ).

Thus Quantum Mechanics gives no rigorous proof. But it gives a certain  hint that
\begin{equation}
(\delta \bar E)_W = 0
\end{equation}
in classical limit. We are going to find rigorous proof of this Variational Principle
in the framework of Classical Mechanics without any reference to Quantum Mechanics.

\section{The Four Variational Principles of Mechanics.}

The starting point is the relation between the Lagrangian $L(q,\dot q)$ and
Hamiltonian $H(q, p)$,
\begin{equation}
L(q,\dot q) = p\dot q -H(q, p) \;\; ,
\end{equation}
which we integrate over some arbitrary time interval $(0,t)$ along
some trial trajectory $q(t'), p(t') $, connecting
fixed endpoints, to obtain
\begin{equation}
S = \int^t_0 dt' L = \int^t_0 dt' p\dot q -\int^t_0 dt' H \;\; .
\end{equation}
Using the definitions of $S$  in eq.(10), $W$  in eq.(4) and $\bar E$  in eq.(3),
we can rewrite eq.(10) as
\begin{equation}
S = W-\bar E t \;\; .
\end{equation}h
For {\it real} trajectories, where $\bar E = E $, equation (11) is
well known [4]. Taking an arbitrary variation of our trial
trajectory $q(t')\to q(t')+\delta q(t')$, with fixed endpoints, as
we assume throughout, and an arbitrary variation of our endtime
$t\to t+\delta t$, we have to first order in $\delta q(t')$ and
$\delta t$:
\begin{equation}
\delta S + \bar E \delta t = \delta W - t\delta \bar E\;\; .
\end{equation}
 We
now show that the two sides of this general kinematic relation
each vanish when we consider variations around a true trajectory,
i.e.
\begin{equation}
\delta S = -E\delta t \; , \;\; \delta W = t\delta\bar E \;\; ,
\end{equation}
where we have used $\bar E = E$  on a true trajectory in
the first of (13).

To derive the first of the relations (13), we recall the Hamilton Principle (HP)
\begin{equation}
(\delta S)_t =0 \;\; ,
\end{equation}
which is valid for  true trajectories. Here  $t$
is fixed.

This relation implies also that for variations around a
true trajectory where both $S$ and $t$ vary we must have
\begin{equation}
\delta S = \lambda\delta t \;\; ,
\end{equation}
where $\lambda$ is some (Lagrange) multiplier.
 Indeed, if (15) were not valid, so that
$\delta S =\lambda\delta t +[$terms dependent on $\delta q(t')]$,
we could have a situation where $\delta t =0$ but
$\delta S \neq 0$ that violates eq. (14). ( We recall that $\delta t$ and $\delta q(t')$ are
independent variations).  In order
to see that $\lambda = -E$, we
specialize to the case of variations between two true trajectories
with endtimes $t$ and $t+dt$. In this case (15) reads $\partial
S/\partial t =\lambda$, and hence $\lambda = -E$ in order to
reproduce the well known relation $\partial S/\partial t = -E$.
This completes the proof of the first, and hence of the second, of
the relations (13).

We now specialize the unconstrained relations (13) by applying
constraints. If, in the first of (13), we take the constraint of
fixed $t$ (i.e. $\delta t =0$), we regain the Hamilton Principle (HP) (14).
 If, on the other hand, we fix $S$, i.e. set $\delta S =0$, we get a
Reciprocal Hamilton Principle  (RHP):
\begin{equation}
(\delta t)_S =0 \;\; .
\end{equation}
This principle  was unknown in the literature in general formulation of eq.(16).
But many special  cases of (RHP) were discussed earlier by many authors starting from
Rayleigh (see  list of references in [2]).

Let us consider the second of the unconstrained relations (13). If we
fix $\bar E$, i.e. set $\delta \bar E = 0$, we get
reformulated Maupertuis' Principle (MP) :
\begin{equation}
(\delta W)_{\bar E} = 0 \;\; .
\end{equation}

In our  reformulation we relax the constraint of fixed energy $E$ for
virtual paths. We allow a larger class of trial trajectories (or "virtual" paths)
which do not necessarily conserve energy $E$. Instead of fixing the energy we
keep the {\it mean energy} $\bar E$ fixed. The old  set of trial trajectories
with fixed $E = \bar E$ is a subset of this larger set. Now  both quantities $W$ and
$\bar E$ are {\it global} whereas in the traditional formulation of MP the constraint  ($E$
fixed) was so to say {\it local}. Thus we get nontrivial generalization of Maupertuis' Principle.

For fixed $W$ (i.e. $\delta W =0$) we get  (RMP) :
\begin{equation}
(\delta\bar E)_W =0 \;\; .
\end{equation}
This is exactly the classical limit of Schr\"{o}dinger's Quantum Variational Principle
that we discussed in the previous section.

This constitutes an abstract proof of the MP and RMP  variational
principles of classical dynamics.
We  have an economical derivation of three other variational
principles starting from the HP. The argument we are using is an adaptation
of Gibbs' familiar argument in thermodynamics [5], when discussing
the Legendre transform relation of free energy, energy,
temperature, and entropy. In the case of Classical Mechanics we have similar
 set of relations for {$S, t, W, \bar E$}:
\begin{equation}
(\delta\bar E)_W =0\stackrel{reciprocity}{\longleftrightarrow} (\delta W)_{\bar E} = 0\;\; ,
\end{equation}

\begin{equation}
(\delta S)_t =0\stackrel{reciprocity}{\longleftrightarrow} (\delta  t)_{S} = 0\;\; ,
\end{equation}

\begin{equation}
(\delta  S)_t =0\stackrel{Legendre}{\longleftrightarrow} (\delta  W)_{\bar E} = 0\;\; .
\end{equation}

It is clear that the four
principles (14), (16), (17) and (18) are equivalent. However,
this does not mean that they are equally useful for solving
particular problems.  The RMP
(18) makes a smooth connection to the Quantum Mechanical
Variational  Principle. We also find that the RMP is well suited
to solve approximately classical problems by a procedure which is
very similar to that used in the variational method of solution of
quantum problems (see [1,2]).

\section{Percival's Variational Principle for Invariant Tori.}

  In this section we are going to consider one interesting application of (RMP).
For integrable systems, all bounded motions are quasi-periodic,
whereas for nonintegrable systems only finite fraction of bounded
motions are quasi-periodic, the rest are chaotic. A quasiperiodic
motion is confined to a torus in phase space. Percival [6] has
derived a variational principle for these tori. We show that his principle
 is a special case of (RMP).

 We consider the initial and final points of the trajectory to be close together,
and the trajectory very long (i.e. the time $T \to \infty $). For a true path, the trajectory
winds around the torus with uniform density, so that time average can be
replaced by phase average:
\begin{equation}
\bar E = \frac{1}{T} \int^T_0 H(q, p)dt =\int\frac{d\Theta^{N}}{(2\pi)^{N}}
H({q}(\Theta_k), {p}(\Theta_k) )\;\; ,
\end{equation}
where ${\Theta_k}$ are the angle variables parametrizing the torus.
The torus is determined by the set of actions $W_k$ that are constants
of motions and that correspond to the set of fundamental loops of
the torus.
 The action $W$ for a long trajectory can be written as
\begin{equation}
  W = \sum_k N_k W_k \;\;.
\end{equation}
\
where $N_k=\nu_{k}T$ is the winding number for angle  $\Theta_k$ and $\nu_k$
is the frequency.

Consider now an arbitrary small deformation of a given invariant
torus and a trial trajectory on the trial torus. We can choose the trial trajectory
such that it covers the trial torus uniformly as well. Thus
\begin{equation}
W=\sum_k N_k \bar W_k \;\;,
\end{equation}
where  $\bar W_k$ is the mean action on the trial torus:
\begin{equation}
\bar W_k = \int\frac{d\Theta^{N-1}}{(2\pi)^{N-1}}W_k = 2\pi\int
\frac{d\Theta^N}{(2\pi)^N}\vec p \cdot \frac{\partial \vec
q}{\partial\Theta_k} \;\; ,
\end{equation}

We are now in position to apply the unconstrained version of the Maupertuis
principle
\begin{equation}
\delta\bar E - \frac {\delta W } {T} = 0 \;\;
\end{equation}
for the considered varied paths. Since the endpoints for
real and trial trajectories are the same, we have equal winding numbers $N_k$
for the two trajectories. Thus equation can be rewritten  as
\begin{equation}
0 = \delta \bar E -\sum_k\left(\frac{N_k}{T}\right) \delta\bar W_k
= \delta\bar E -\sum_k \nu_k \delta\bar W_k
\end{equation}
or
\begin{equation}
\delta (\bar E -\sum_k\nu_k \bar W_k) = 0 \;\;,
\end{equation}
where $\nu_k$ are the frequencies for the invariant torus. This equation is
Percival's variational principle for invariant tori [6]: the mean energy $\bar E$
is extremized with the set of mean actions $\bar W_k$ held fixed. The constant
 Lagrange multipliers $\nu_k$ are the invariant torus frequencies.
\begin{equation}
(\delta \bar E )_{{\bar W_k}} = 0 \;\;.
\end{equation}
  In the time of Old Quantum
Mechanics this relation was used as a postulate for quantization of integrable systems [7].


\begin{thebibliography}{99}
\bibitem{1}
G.Karl and V.A. Novikov, Phys. Rev. {\bf D51} (1995) 5069; J. Exp.
Theor. Phys.{\bf 80} (1995) 783.
\bibitem{1}
C.G.Gray,G.Karl and V.A. Novikov, Ann.Phys.{\bf 251} (1996) 1;
Am.J.Phys. {\bf 64} (1996) 1172; Am.J.Phys. {\bf 67} (1999) 959.
\bibitem{3}
R.P.Feynman, Rev.Mod.Phys.{\bf 20} (1948) 267; see also
P.A.M.Dirac, Physik.Z.Sowjetunion {\bf 3} (1933) 64.
\bibitem{4}
See, e.g. L.D. Landau and E.M. Lifshitz, {\it Mechanics}, 2nd ed.,
p. 26, Addison-Wesley, Reading, MA, 1969.
\bibitem{5}
J.W.Gibbs, "The Scientific Papers of J. Willard Gibbs", Vol.1, "Thermodynamics",
p.56, Longmass, Green, London, 1906: reprinted by Dover, New York, 1961.
\bibitem{6}
I.C. Percival, J. Phys. A: Math. Nucl. Gen. {\bf 7} (1974) 794.
\bibitem{7}
J.H. Van Vleck, Phys. Rev. {\bf 22} (1923) 547 .
\end{thebibliography}
\end{document}